\documentclass[]{aa} 
\begin{document}
\outer\def\gtae {$\buildrel {\lower3pt\hbox{$>$}} \over 
{\lower2pt\hbox{$\sim$}} $}
\outer\def\ltae {$\buildrel {\lower3pt\hbox{$<$}} \over 
{\lower2pt\hbox{$\sim$}} $}
\newcommand{\ergscm} {erg cm$^{-2}$s$^{-1}$}
\newcommand{\ergss} {erg~s$^{-1}$}
\newcommand{\ergsd} {erg~s$^{-1}$ $d^{2}_{100}$}
\newcommand{\pcmsq} {cm$^{-2}$}
\newcommand{\ros} {\sl ROSAT}
\newcommand{\exo} {\sl EXOSAT}
\newcommand{\xmm} {\sl XMM-Newton}
\newcommand{\chan} {\sl Chandra}
\def\rchi{{${\chi}_{\nu}^{2}$}}
\def\uchi{{${\chi}^{2}$}}
\newcommand{\Msun} {$M_{\odot}$}
\newcommand{\Mwd} {$M_{\rm wd}$}
\def\Mdot{\hbox{$\dot M$}}
\def\mdot{\hbox{$\dot m$}}
\input psfig.sty

\title{A search for stellar X-ray sources 
   in the Sagittarius and Carina dwarf
galaxies - I: X-ray observations} 
\titlerunning{X-ray observations of the Sgr and Car dSph}
\authorrunning{Ramsay \& Wu}

\author{Gavin Ramsay\inst{1} and Kinwah Wu\inst{1,2}}

\offprints{G. Ramsay: gtbr@mssl.ucl.ac.uk}

\institute{
$^{1}$ Mullard Space Science Laboratory, University College London,
Holmbury St. Mary, Dorking, Surrey, RH5 6NT, UK \\ 
$^{2}$ School of Physics A28, University of Sydney, NSW 2006, Australia}

\date{Accepted: 26th August 2006}

\abstract{ {\sl ROSAT} observations found no convincing evidence for
X-ray sources located in local dwarf spheroidal galaxies (dSph).  Now
with more sensitive instruments on board {\sl Chandra} and {\xmm} we
can reach fainter luminosity levels. We report on an observation of
the Sagittarius (Sgr) dSph made using {\sl Chandra} and an observation
of the Carina (Car) dSph made using {\xmm}.  Our observations are
sensitive to sources with X-ray luminosities in the $0.1-10$~keV band
of $\sim1\times10^{32}$ \ergss and 3$\times10^{34}$ \ergss for the Sgr
and Car fields respectively.  We have identified a total of 80 sources
in the Sgr field and 53 sources in the Car field.  Although the source
numbers are roughly consistent with the expected number of background
AGN, we found a small fraction of X-ray sources which were soft and
could be located in the host dSph.  Follow-up optical/IR observations
may help to identify their optical counterparts and hence determine
their nature.  
}

\maketitle

\section{Introduction}

Dwarf galaxies are low luminosity galaxies and make up a large
fraction of galaxies at the present epoch (Marzke \& da Costa 1997).
Because of their faintness, the best studied systems reside in the
Local Group.  The Milky Way has 12 known satellite galaxies.  With the
exception of the LMC and SMC, which are irregulars, all of them are
dwarf spheroidals (dSphs).  DSphs are generally gas poor and have low
metallicity.  Their star formation activity has long since ceased, and
their stellar populations are old.  (See Mateo 1998 for a review of
Local Group dwarfs.)

While the Local Group dSphs have been well studied in the optical and
infra-red, their X-ray properties are yet to be established.  DSphs
were seldom observed in the X-ray band, in the belief that they do not
contain substantial number of detectable X-ray sources.  A {\sl ROSAT}
observation of the Fornax dSph (which is at a distance of 138 kpc)
found 19 sources in the field (Gizis, Mould \& Djorgovski 1993).
Since this number was consistent with the expected number of
background AGN, the sources were not thought to be stellar binary
sources located in the dSph.  A recent {\sl Chandra} observation of
the Sculptor dSph (at a distance of 79 kpc) identified 5 sources which
were likely to be located in that galaxy (Maccarone et al.\
2005). They have inferred X-ray luminosities $L_{\rm x} \sim 6-90
\times 10^{33}$~\ergss, which are well below $10^{37}-10^{38}$~\ergss,
typical luminosities of bright Galactic X-ray binaries.  From these
studies we may conclude that dSphs in the Local Group do not have a
significant number of bright X-ray sources (see Zang \& Meurs 2001).
 
The number of binary X-ray sources in a galaxy is determined by the
formation rate of the sources and the rate at which the sources cease
to be X-ray active.  Also, it depends on the duration of the X-ray
duty cycles of the sources.  Observations show that X-ray sources are
more abundant in starburst galaxies and interacting galaxies than in
the `milder' galaxies (see e.g.\ Fabbiano 2003), and the brightness
distribution of X-ray sources vary with their host environments even
within the same galaxy.  For instance in spiral galaxies, e.g. M81
(Tennant et al.\ 2001; Swartz et al.\ 2003), sources in the galactic
disks have power-law luminosity functions, while sources in galaxy
bulges tend to have broken power-law like or exponential cut-off type
luminosity functions.  Now there is strong evidence that X-ray source
populations in galaxies and their brightness distribution are
dependent on the star formation activity of the source host
environment in recent epochs.  This phenomenon can be explained by a
birth-death process as that described in Wu (2001).  The lack of very
bright sources in old stellar populations can be explained by the fact
that X-ray sources have finite life-spans and bright sources which
require rapid mass transfer are short-lived.  If the mass transfer
rate in a binary is sufficiently low or if the mass-transfer active
phase in its duty cycle is sufficiently short, the system may survive
a very long time as an X-ray source.  The source may exhibit
alternating active and faint states, and so it may not be easily
identified in a single snap-shot observation.

In the birth-death model presented in Wu (2001), there is also an old
X-ray source population in any galaxy whose progenitors were the
first-generation stars formed in the very early epoch.  These
`primordial' binary X-ray sources are dim, and their mass-donor stars
are low-mass stars.  As these binaries have a sustainable
mass-transfer phase spanning over a period of $> 10$~Gyr, the average
mass-transfer rate between cannot exceed $10^{15}$~g~s$^{-1}$.  If the
accreting compact object is a black hole or a neutron star, the X-ray
luminosity of the binary would be $< 10^{35}$~\ergss.  Even when a
detection limit of well below $10^{35}$~\ergss has been achieved, the
old binary X-ray sources are not easily identified in big spiral
galaxies like M31 or M81 (see e.g.\ Kong et al.\ 2003; Swartz et al.\
2003), because there are hundreds of younger X-ray sources with
luminosities above $10^{36}$~\ergss.  DSphs have not had recent star
formation activity and are therefore the most appropriate targets to
search for the old X-ray source population.

{\sl ROSAT} had the sensitivity to detect sources brighter than $\sim
10^{34}$~\ergss \ in the nearby dSphs but were unable to detect
sources with luminosities of $10^{32}$~\ergss.  With {\it Chandra} and
{\sl XMM-Newton}, we can now search for sources with luminosities
below $10^{32}$~\ergss\ in the dwarf satellites of the Milky Way.  We
have made X-ray observations of the Sagittarius dSph (Sgr) with {\sl
Chandra} in 2003 and Carina dSph (Car) with {\sl XMM-Newton} in 2004.
This paper presents the results of our X-ray observations of these two
galaxies.  (The {\sl Chandra} observation of Sgr was centered near the
globular cluster M54, and analysis of the cluster sources has already
been presented in Ramsay \& Wu 2006).

Sgr, at a distance of 27.4~kpc (Layden \& Sarajedini 2000), is the
second closest satellite to the Milky Way.  It extends over at least
$22^{\circ} \times 8^{\circ}$ in the sky, and is located on the far
side of the Milky Way to the Sun (Ibata, Gilmore \& Irwin 1994, 1995).
The absorption column density along the line-of-sight is relatively
high ($1.2 \times 10^{21}$~\pcmsq) because of the low galactic
latitude of Sgr ($-14^{\circ}$), Star formation in Sgr mostly ceased
$\sim$10 Gyr ago (Mateo 1998).  Car was discovered in plates taken
using the UK Schmidt Telescope (Cannon, Hawarden \& Tritton 1977) and
is 100 kpc distant (van den Bergh 2000).  The line-of-sight absorption
column to Car is $3.9 \times10^{20}$~\pcmsq.  Car has had a more
complex star formation history than Sgr: around half of its stars were
formed 7 Gyr ago, the remainder occuring at 15 and 3 Gyr ago
(Hurley-Keller, Mateo \& Nemec 1998).

\section{Observations and Data Analysis}

\subsection{Sagittarius Dwarf Galaxy}

Sgr was observed with {\sl Chandra} ACIS-I for 30~ksec on 2003 Sep 3.
The total field of view was $16.9{'} \times16.9{'}$.  The pointing was
centered on $\alpha_{2000}=18^{h} 55^{m} 3.0^{s}$, $\delta_{2000} =
-30^{\circ} 28{'} 59{''}$: this was offset from the center of the
globular cluster M54 by $-12{''}$ in declination.  Front illuminated
CCDs $0-3$ were used.  The `read' mode was configured in {\tt TIMING}
mode; the `data' mode was {\tt VERY FAINT}.  During the observation
the solar particle background was very low.

We used the primary data products as supplied by the {\chan} Data
Archive Operations.  A source search was initially performed on each
CCD separately in the $0.3-8.0$~keV band using {\tt wavedetect} in the
CIAO v3.0 software package.  This was done in conjunction with an
exposure map using a mean photon energy of 1.5~keV.  We rejected all
`sources' with fewer than 5 counts (not background subtracted) in the
$0.3-8$~keV band.  The resulting source list was then compared by-eye
with the image in that band: `sources' which were not clear in the
image were not counted.  We detected a total of 80 sources, excluding
the 7 sources belonging to the globular cluster M54 (Ramsay \& Wu
2006).  Assuming a power law spectral model with a spectral index of
1.4 and an absorption column of $1.2 \times 10^{21}$~\pcmsq, we obtain
a sensitivity level of $4.0 \times10^{-15}$~erg~cm$^{-2}$s$^{-1}$
(unabsorbed flux) in the $0.3-8$~keV band.  The inferred unabsorbed
flux in the $0.1-10$~keV band is $4.6
\times10^{-15}$~erg~cm$^{-2}$s$^{-1}$, corresponding to a luminosity
of $1.1 \times 10^{32}$~\ergss\ for the distance to Sgr.

Using the source list derived above, we also determined the counts for
each source in the $0.3-1$~keV (S), $1-2$~keV (M) and $2-8$~keV (H)
bands (which will be used to construct the colour-colour plots
following the convention considered in Prestwich et al.\ (2003), Soria
\& Wu (2003) and Swartz et al.\ (2004)).  To determine the mean
background count rate in each band we created an event file which
excluded the detected sources.  We then scaled the number of events in
the sampled area with the mean size of the extraction region. In the
$2-8$~keV band a total of 45 sources are detected above the
2.0-$\sigma$ level.  We show the list of sources with their counts in
each band in the Appendix.

\subsection{Carina Dwarf Galaxy}

Car was observed by {\xmm} on 2005 May 25 for 41.7~ksec.  Both the
EPIC MOS (Turner et al.\ 2001) and the EPIC pn (Str\"{u}der et al.\
2001) wide field instruments were in full frame mode and used thin
filters.  The diameter of the EPIC MOS field is approximately $28{'}$,
and the field of view of the EPIC pn is approximately
$27{'}\times27{'}$.  The field center was $\alpha_{2000}= 6^{h} 41^{m}
37^{s}$, $\delta_{2000}=-50^{\circ} 58{'} 0{''}$.  The data were
analysed using the {\sl Science Analysis Software} (SAS) v6.0.

The calibrated MOS1 and MOS2 event data were combined and analysed
separately from the EPIC pn data. Data from the first 400~sec were
excluded from the analysis since the background was significantly
higher than the following data. Events which had {\tt PATTERN=0-12}
and {\tt FLAG=0} were used. Images and exposure maps were extracted.

These data were searched for sources with the resulting source list
being examined by eye - `sources' which were not visible as such in
the $0.2-10$~keV images were removed from the source list. Some
sources were only detected in the MOS instruments because they were
located in the pn chip gaps (and vice-versa).  A total of 53 sources
were detected.  Their count rates were determined in the energy
ranges: $0.2-1$~keV, $1-2$~keV and $2-10$~keV.  In the $2-10$~keV band
41 sources were detected above 2.0-$\sigma$ level.  The source
positions, count rates and X-ray colours are given in the Appendix.

Assuming a power law spectral model with a spectral index of 1.4 and
an absorption column of 3.9$\times10^{20}$ \pcmsq, we obtain a flux
sensitivity level of $2.8 \times 10^{-14}$~erg~cm$^{-2}$s$^{-1}$ in
the $0.2-10$~keV band (unabsorbed).  This corresponds to an unabsorbed
flux of $2.9 \times 10^{-14}$~erg~cm$^{-2}$s$^{-1}$ in the
$0.1-10$~keV band, and a luminosity of 2.9$\times10^{34}$~\ergss\ for
the distance to Car.

\section{Source brightness distributions} 
\label{distribution}

\begin{figure}
\begin{center}
\setlength{\unitlength}{1cm}
\begin{picture}(6,5.5)
\put(-1.5,-0.){\includegraphics{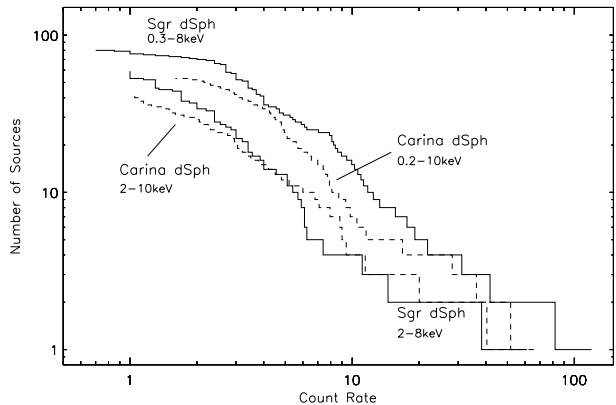}}
\end{picture}
\end{center}
\caption{The cumulative luminosity functions ($N(>S)$ as a function of
count rate $S$) of the sources in the Sgr and Car fields in 2 energy
bands.  The count rate for the Sgr sources is ct/10ksec and ct/ksec
for the Car sources.}
\label{cumulative}
\end{figure}

Figure \ref{cumulative} shows the cumulative luminosity function, $N
(>S)$ (where $S$ is the count rate), of the Sgr and Car sources.  The
luminosity function of the Sgr sources in the $0.3-8$~keV band shows a
flattening at its faint end where the count rate falls below $\sim 2.3
\times 10^{-4}$~ct~s$^{-1}$.  The flattening is not seen in the
luminosity function for the $2-8$~keV band.  The luminosity functions
of the Car sources in the $0.2-10$~keV and $2-10$~keV bands appear
similar and have no obvious features.  The flattening in the
luminosity function of Sgr sources at the $0.3-8$~keV band is probably
not due to the change in the distribution of the intrinsic brightness
of the sources but incompleteness caused by high absorption at
energies below 1~keV in the direction to galaxy.

We used the Bayesian algorithm, derived by Wheatland (2004), to
determine the slope of the luminosity function and obtained a
power-law index $\alpha = 1.42\pm0.16$ for the Sgr sources with a
cut-off at 2.3 counts/10ksec at the $0.3-8$~keV band.  The slope is
slightly steeper than that for the sources in the globular cluster M54
($0.90\pm0.34$, Ramsay \& Wu 2006).  For the luminosity function of
the Car sources in the $0.2-10$~keV band we obtained a power-law slope
of $\alpha=1.39\pm0.05$.  At the $2-8$~keV band the slope of the Sgr
sources (1.03$\pm$0.13) is flatter than that of the Car sources
(1.36$\pm$0.06).  They are steeper than the slope of the sources
($0.61 \pm 0.10$) in the {\sl Chandra} north and south deep fields in
the $2-10$~keV band (Rosati et al.\ 2002).  However, they are similar
to the slope in the $5-10$~keV band (1.35$\pm$0.15).

To determine if any of the sources were bright foreground stars
we correlated our X-ray positions with the the UCAC2 catalogue
(Zacharias 2004). This catalogue is complete down to $R$=16 and has
astrometry better than 70 milli-arcsec. For the Sgr field we found 5
X-ray sources which had optical counterparts within a 2 arcsec radius,
while for Car, we found 6 sources within a 4 arcsec radius (the error
circle for the {\xmm} positions were greater than for {\chan}). The
UCAC2 mag for these sources is given in Tables A1--A3. We classify
these sources as likely foreground objects.

We estimated the expected number of background AGN in the field of
view using the luminosity function of the {\sl Chandra} Deep Field
South (CDF-S) sources.  We considered the $2-10$~keV band in which the
effect of extinction is reduced.  Assuming a power-law spectral model
with a spectral index of 1.4 and an absorption column density to the
edge of our galaxy in the line-of-sight, the sensitivity limit in the
$2-10$~keV band is $3.7 \times10^{-15}$ \ergscm\ for the Sgr sources
and 1.9$\times10^{-14}$ \ergscm\ for the Car sources.  The number of
background AGN in the Sgr field would be about $55-79$ and the number
in the Car field about $16-34$.  The number of Sgr sources in the
2--8keV band is 42 (after the subtraction of 3 sources which are
likely foreground objects), which is slightly below the expected
number if background AGN.  The number of identified sources in the
2--10keV band in the Car field is 40 (after the subraction of 1 source
which was a likely foreground object), which is slightly above the
expected number of background AGN.  Whilst it is likely that the
majority of the sources in the fields are background AGN, we cannot
rule out at this stage the possibility that a few of the sources are
stellar binary sources belonging to the two dSphs.

\section{Luminosity estimates}
\label{luminosity}      

In \S \ref{distribution} we estimated the sensitivity limits in
various energy bands assuming a power law spectral model of slope 1.4.
We also estimated the count to flux conversions assuming a power-law
spectral model with slope of 1.7 and with a blackbody model with
$kT_{\rm bb}=0.1$~keV and found that making an assumption of one
conversion factor for all spectra -- irrespective of their colour --
in general gives an uncertainty of 30\% in the source luminosity.

\subsection{Sagittarius Dwarf Galaxy}

We applied a conversion in which $10^{-4}$~ct~s$^{-1}$ in the
$0.3-8$~keV band is equivalent to an unabsorbed flux of
1.8$\times10^{-15}$ \ergscm\ in the $0.1-10$~keV band.  For this
conversion, the inferred luminosities of the Sgr sources would be
$\sim1\times10^{32}-2\times10^{34}$ \ergss, if they are associated
with the galaxy.  The luminosities are similar to those of cataclysmic
variables (which are interacting binaries containing a low-mass star
transferring mass to a white dwarf) in their active state but are
below the range for Galactic X-ray binaries in their bright X-ray
states.
   
\subsection{Carina Dwarf  Galaxy}

We applied a conversion in which 1 ct/ksec in the $0.2-10$~keV band is
equivalent to an unabsorbed flux of 1.6$\times10^{-14}$ \ergscm\ in
the $0.1-10$~keV band.  If the sources are in Car, they would have
luminosities $\sim 3\times10^{34}-1\times10^{36}$ \ergss.  These
luminosities are higher than those of cataclysmic variables but are
comparable with those of X-ray binaries powered by wind-fed accretion
(cf.\ sources in Table 1 of Muno et al.\ 2003).

\section{X-ray colours}
\label{spec}  

\subsection{Sagittarius Dwarf Galaxy} 

\begin{figure}
\begin{center}
\setlength{\unitlength}{1cm}
\begin{picture}(6,6.5)
\put(-1.5,-3.8){\includegraphics{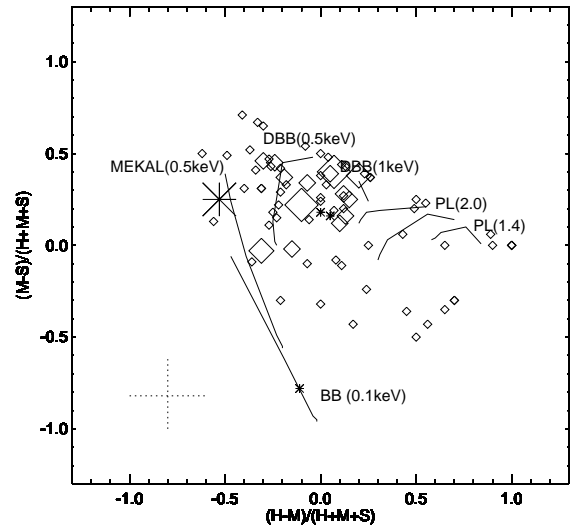}}
\end{picture}
\end{center}
\caption{The colour-colour plot for the sources in the Sgr field.  S,
M and H correspond to photon count rates in the $0.3-1$, $1-2$ and
$2-8$~keV bands.  The size of the symbol indicates the number of
photon counts in the $0.3 - 8$~keV band. The solid lines show the
predicted colour-colour track assuming different models: BB
(Blackbody), MEKAL (optically thin plasma), DBB (disk blackbody), PL
(power law). The temperature/slope of the model is shown in
brackets. The absorption is assumed to start at the absorption to the
edge of our Galaxy and then increases up to $N_{H}=5\times10^{22}$
\pcmsq. In the lower corner we show the typical size of the errorbars
for a moderately bright source (source number 31). Sources which had
bright optical counterparts (and hence likely foreground stars) are shown
as `stars'.}
\label{sgr_colour}
\end{figure}

Figure \ref{sgr_colour} shows the colour-colour plots of the sources
in the Sgr field.  We considered the colours x$\equiv(H-M)/(H+M+S)$
and y$\equiv(M-S)/(H+M+S)$ following the convention of Prestwich et
al.\ (2003), Soria \& Wu (2003) and Swartz et al.\ (2004) that S, M
and H corresponding to count rates in the $0.3-1$, $1-2$ and $2-8$~keV
bands.  We compared the colours of the sources with 4 spectral models:
a blackbody model ($kT_{\rm bb} =0.1$~keV), a disk blackbody model
($kT_{\rm dbb} = 0.5$ and 1~keV), an optically thin thermal plasma
model ($kT_{\rm th} = 0.5$~keV) and a power-law model ($\Gamma=1.4$
and 2.0).  In each model, we considered a range of absorption column
densities ranging from the line-of-sight Galactic absorption ($N_{\rm
H}=1.2\times10^{21}$~cm$^{-2}$) to a hydrogen column density of
$N_{\rm H} = 5\times 10^{22}$~cm$^{-2}$.  The colour tracks of the
models generally start at low y-colour values and then increase with
absorption.  Except for the very soft spectral models (e.g. blackbody
with $kT_{\rm bb} = 0.1$~keV) the x-colour value increases with
absorption.  For sufficiently large absorption the colour tracks of
the hard models (e.g. power law models, regardless of the photon
index) converge.  The soft models with different parameters, however,
follow separate tracks.

The majority of the sources in the Sgr field have colours consistent
with those of the absorbed power law or absorbed disk blackbody
spectra, although some sources may require combined blackbody and
power-law spectra to explain their colours.  These spectra are typical
of AGN and Galactic X-ray binaries, but we can single out the few
stellar candidates from the background sources easily based on this
set of X-ray data alone.  We note that some of the Sgr sources (No. 6,
9, 30, 46, 57 and 69 in Tables A1 -- A2 ) have very soft spectra.
They lie close to the tracks of the blackbody ($kT_{\rm bb} =0.1$~keV)
model and the optically thin thermal ($kT_{\rm th} = 0.5$~keV) model.
Their colours are not consistent with AGN X-ray spectra, which would
lie closer to the harder colour-colour tracks.
  
\subsection{Carina  Dwarf Galaxy} 

\begin{figure}
\begin{center}
\setlength{\unitlength}{1cm}
\begin{picture}(6,6.5)
\put(-1.5,-3.8){\includegraphics{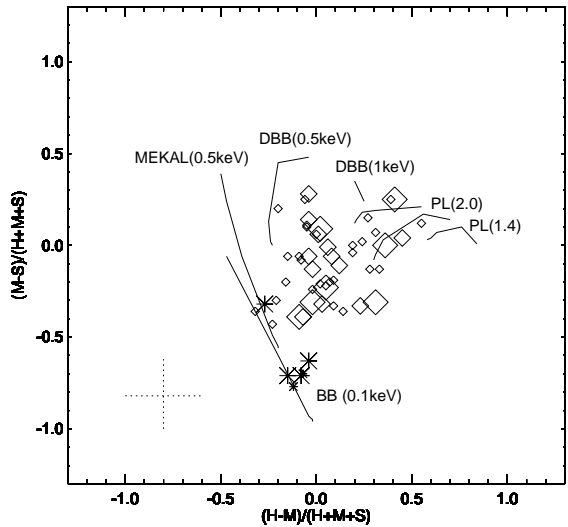}}
\end{picture}
\end{center}
\caption{As Figure 2 but for sources in the Car field. Otherwise S, M
and H correspond to photon count rates in the $0.2-1$, $1-2$ and
$2-10$~keV bands.  The size of the symbol indicates the number of
photon counts in the $0.2-10$~keV band. The absorption is assumed to
start at the absorption to the edge of our Galaxy and then increases
up to $N_{H}=1\times10^{22}$ \pcmsq. In the lower corner we show the
typical size of the errorbars for a moderately bright source (source
number 23).}
\label{carina_colour}
\end{figure}

Figure \ref{carina_colour} shows the colours of the Car sources in the
(H-M)/(H+M+S), (M-S)/(H+M+S) plane.  Here S, M and H represent the
$0.2-1$, $1-2$ and $2-10$~keV bands.  The colours for various spectral
models with absorption column densities starting from the column
density to the edge of the Galaxy (3.9$\times10^{20}$~\pcmsq) to the
column density of $1 \times10^{22}$~\pcmsq are also shown for
comparison.  Despite the different detectors and slightly different
energy bands used, the model tracks are similar to those in the
colour-colour plot of the Sgr case.  The sources have a wide spread in
colour, as expected for background AGN with different absorption
column density.  There are roughly 3 sources which lie close to the
tracks of the blackbody model ($kT_{\rm bb} = 0.1$~keV) and optically
thin thermal plasma model ($kT_{\rm th} = 0.5$~keV).  They are sources
No. 17, 25 and 30 as listed in Table A3.  The colours of these sources
are not consistent with AGN, and they are likely to be foreground
stellar sources or sources inside the Car dSph.

\section{Discussion \& Conclusions}

We have performed X-ray observations of the Sgr and Car fields.  Our
observations reached a sensitivity limit (unabsorbed) $\sim
1\times10^{32}$ \ergss\ for the Sgr field and $3\times10^{34}$ \ergss
for the Car field.  We have identified a total of 80 sources (in the
$0.3-8$~keV energy band) in the Sgr field (excluding the 7 sources
which we have identified as being in the globular cluster M54) and 53
sources (in the $0.2-10$~keV band) in the Car field. At energies
$>$2keV we found 42 sources in the Sgr field and 40 in the Car field
(these numbers do not include objects which are likely to be
foreground stars).  Based on the CDF-S we estimate that $55-79$
background AGN would be in the Sgr field and $16-34$ in the Car field.
(The expected numbers will vary to some extent since the number of AGN
varies from field to field.)  The source numbers in the two field are
broadly consistent with the expected number of background AGN.
However, from the analysis of X-ray colours, we have found a small
number of sources which have colours which are difficult to be
explained by spectral models appropriates for AGN. We therefore
propose that these are candidate stellar X-ray sources. The nature of
these sources may be determined by follow-up X-ray observations or by
identifying their optical/IR counterparts. Both Sgr and Car have low
galactic latitudes ($b=-14^{\circ}$ and $-22^{\circ}$ respectively),
which gives rise to a number of possible optical counterparts for many
of the X-ray sources. The search for the optical counterparts to the
X-ray sources will be discuss in detail in a future paper.

We note that Maccarone et al.\ (2005) found a small number of X-ray
sources which are believed to belong to the Sculptor (Scu) dSph.
These sources have X-ray luminosities $L_{X} = 6 -
90\times10^{33}$\ergss.  The optical counterparts of these sources are
relatively easy to identify as Scu has a high galactic latitude
($b=-83^{\circ}$).  The interstellar absorption to the edge of the
Galaxy towards the direction of Car and Scu are similar
(3.9$\times10^{20}$ \pcmsq and 2.0$\times10^{20}$ \pcmsq
respectively).  If we take a naive view that the number of X-ray
sources are proportional to total stellar mass of the host galaxy,
then we may obtain an estimate of the number of sources to be detected
in Car from the results of Scu (we acknowledge that the reality is likely 
to be more complex since they have different star formation histories).

The Scu dSph has a core radius of 5.8$^{'}$ and the field of view of
the {\chan} observations of Scu covered an area of
$\sim17^{'}\times17^{'}$.  The observations therefore covered a large
fraction of the galaxies mass (6.4$\times10^{6}$\Msun).  The field of
view of the {\xmm} observations of Car were much larger than the core
radius of 8.8$^{'}$ and covered a large fraction of the galaxies mass
(1.3$\times10^{7}$\Msun).  (These parameters are taken from Mateo
1998). The {\xmm} observations of Car were sensitive to X-ray
luminosities $>3\times10^{34}$ \ergss: there were 2 Scu sources with
luminosities greater than this.  If we assume that the number of X-ray
sources scales with the total stellar mass of the galaxy there will be
roughly 2 sources with $L_{X}>3\times10^{34}$ \ergss\ in Car.  This
suggests that some of the 3 soft X-ray sources in the Car field could
actually be located in the Car dSph.

We observed only a small fraction of the sky coverage of the Sgr dSph
-- roughly 0.05\%.  The total mass of the Sgr dSph is $\sim10^{9}$
\Msun (Ibata Gilmore \& Irwin 1995), but we sampled a fraction of the
galaxy with a total mass of $\sim5\times10^{5}$ \Msun.  If we scale
the number of X-ray sources with mass, 1 X-ray source in Sgr would
imply a total number $4 \times10^{5}$ of X-ray sources with $L_{X} >
10^{32}$ \ergss\ in the Milky Way. This number is obviously
rather uncertain for various reasons, and what fraction of these
sources can be detected in snap-shot is unclear given that a
substantial fraction of the sources are transients or variable
sources.

Birth-death models (e.g.\ Wu 2001) predict a primordial X-ray source
population in all galaxies.  These sources are remnants of the first
generation stars formed with the galaxy. Suppose that the slope of the
luminosity function of these source is similar to the faint end of the
luminosity function of the bright bulge sources in M81 (see Tennant et
al.\ 2001), we would set an upper limit that fewer than 100 primordial
X-ray sources with luminosity exceeding $\sim 10^{36}$\ergss\ are in
large spiral galaxies similar to M81 or the Milky Way.  However, the
luminosity function of the primordial X-ray sources would show an
exponential cut-off (a consequence of aging effects), beyond which the
source counts drop rapidly.  The number of primordial sources with
such luminosities should in fact be much fewer since fainter sources
can survive for a long time.  Our observation of Sgr was sensitive
down to $L_{X}\sim 1 \times 10^{32}$\ergss, which is sufficient to
detect most of the persistent bright X-ray binaries and moderately
bright CVs.  If the soft sources which we found belong to the host
dSph, then they may be candidate primordial X-ray sources.

\section{Acknowledgments}

This is work based on observations obtained with {\sl XMM-Newton}, an
ESA science mission with instruments and contributions directly funded
by ESA Member States and the USA (NASA). We thank Joe DePasquale for
help in scheduling the {\chan} observations. KW acknowledges the
support for his visits to the University of Sydney through an
Australian NSW State Expatriate Researcher Award. We thank the
anonymous referee for useful comments and suggestions.

\appendix

\section{Source lists}

\renewcommand{\baselinestretch}{1.0}

\begin{table*}
\begin{center}
\begin{tabular}{l@{\hspace{2pt}}c@{\hspace{5pt}}crrrrrrrrrrrr@{\hspace{2pt}}r}
\hline
Src & RA & Dec & \multicolumn{2}{r}{0.3-8keV} & \multicolumn{2}{r}{0.3-1keV} & 
\multicolumn{2}{r}{1-2keV} & \multicolumn{2}{r}{2-8keV} & \multicolumn{2}{c}{HR1} & \multicolumn{2}{c}{HR2} &UCAC2 \\
    & \multicolumn{2}{c}{(2000)} & & $\pm$ &  & $\pm$ & & $\pm$ & & $\pm$ & & $\pm$  & & $\pm$ & mag\\
\hline
 1&   18  54 26.24&  -30  29  40.3&   12.1&   2.0&   1.3&   0.7&  4.7 &   1.2&  6.1  &  1.4   &0.12  &  0.17  &  0.28  &  0.20&\\
 2&   18  54 26.50&  -30  34  24.3&    2.7&   0.9&   0.3&   0.3&  1.0 &   0.6&  1.3  &  0.7   &0.12  &  0.36  &  0.27  &  0.31&\\
 3&   18  54 26.67&  -30  27  08.1&    3.0&   1.0&   0.3&   0.3&  1.3 &   0.7&  1.4  &  0.7   &0.03  &  0.33  &  0.33  &  0.32&\\
 4&   18  54 27.43&  -30  31  52.9&    1.0&   0.6&   0.0&   0.0&  0.0 &   0.0&  1.0  &  0.6   &1.00  &  1.16  &  0.00  &  0.00&\\
 5&   18  54 28.14&  -30  33  30.7&    1.7&   0.7&   0.3&   0.3&  0.7 &   0.5&  0.7  &  0.5   &0.00  &  0.00  &  0.24  &  0.37&\\
 6&   18  54 33.14&  -30  31  21.5&    3.4&   1.1&   2.0&   0.8&  1.0 &   0.6&  0.3  &  0.3   &-0.21  &  0.24  & -0.30  &  0.36&\\
 7&   18  54 33.32&  -30  27  49.9&    7.7&   1.6&   2.0&   0.8&  4.4 &   1.2&  1.3  &  0.7   &-0.40  &  0.30  &  0.31  &  0.26&\\
 8&   18  54 34.05&  -30  25  06.3&   10.8&   1.9&   2.0&   0.8&  3.7 &   1.1&  5.1  &  1.3   &0.13  &  0.18  &  0.16  &  0.16&\\
 9&   18  54 35.25&  -30  26  29.3&    4.0&   1.2&   0.7&   0.5&  2.7 &   0.9&  0.7  &  0.5   &-0.49  &  0.39  &  0.49  &  0.39&\\
10&   18  54 35.38&  -30  31  28.1&    3.4&   1.1&   0.7&   0.5&  1.7 &   0.7&  1.0  &  0.6   &-0.21  &  0.30  &  0.29  &  0.31&\\
11&   18  54 36.00&  -30  25  10.4&    8.1&   1.6&   1.0&   0.6&  4.4 &   1.2&  2.7  &  0.9   &-0.21  &  0.22  &  0.42  &  0.30&\\
12&   18  54 36.40&  -30  27  27.9&    1.0&   0.6&   0.3&   0.3&  0.0 &   0.0&  0.7  &  0.5   &0.70  &  0.71  & -0.30  &  0.37&\\
13&   18  54 36.41&  -30  27  40.7&    3.7&   1.1&   1.3&   0.7&  1.7 &   0.7&  0.7  &  0.5   &-0.27  &  0.28  &  0.11  &  0.27&\\
14&   18  54 37.35&  -30  34  23.8&    2.7&   0.9&   1.0&   0.6&  0.7 &   0.5&  1.0  &  0.6   &0.11  &  0.30  & -0.11  &  0.30&\\
15&   18  54 37.66&  -30  32  25.7&    5.1&   1.3&   1.3&   0.7&  2.0 &   0.8&  1.7  &  0.7   &-0.06  &  0.22  &  0.14  &  0.23&\\
16&   18  54 38.93&  -30  31  29.3&    4.0&   1.2&   0.7&   0.5&  2.4 &   0.9&  1.0  &  0.6   &-0.34  &  0.33  &  0.41  &  0.35&\\
17&   18  54 39.08&  -30  26  08.7&    8.4&   1.7&   0.3&   0.3&  2.0 &   0.8&  6.1  &  1.4   &0.49  &  0.37  &  0.20  &  0.17&\\
18&   18  54 40.17&  -30  27  01.6&    3.0&   1.0&   1.7&   0.7&  0.7 &   0.5&  0.7  &  0.5   &0.00  &  0.00  & -0.32  &  0.34&\\
19&   18  54 40.59&  -30  32  47.7&    1.7&   0.7&   0.7&   0.5&  0.3 &   0.3&  0.7  &  0.5   &0.24  &  0.37  & -0.24  &  0.37&\\
20&   18  54 41.54&  -30  27  44.4&    4.0&   1.2&   1.7&   0.7&  1.3 &   0.7&  1.0  &  0.6   &-0.07  &  0.23  & -0.10  &  0.25&\\
21&   18  54 42.13&  -30  36  51.1&    4.7&   1.2&   2.0&   0.8&  0.3 &   0.3&  2.4  &  0.9   &0.45  &  0.34  & -0.36  &  0.29&\\
22&   18  54 43.00&  -30  34  11.6&    1.3&   0.7&   0.3&   0.3&  0.7 &   0.5&  0.3  &  0.3   &-0.31  &  0.48  &  0.31  &  0.48&\\
23&   18  54 43.67&  -30  26  35.4&    6.4&   1.5&   0.7&   0.5&  3.7 &   1.1&  2.0  &  0.8   &-0.27  &  0.27  &  0.47  &  0.34&\\
24&   18  54 43.89&  -30  29  40.4&    2.4&   0.9&   1.0&   0.6&  0.0 &   0.0&  1.3  &  0.7   &0.56  &  0.50  & -0.43  &  0.40&\\
25&   18  54 44.49&  -30  25  09.5&    4.0&   1.2&   1.0&   0.6&  1.0 &   0.6&  2.0  &  0.8   &0.25  &  0.29  &  0.00  &  0.00&\\
26&   18  54 45.51&  -30  25  11.3&   10.4&   1.9&   1.3&   0.7&  6.1 &   1.4&  3.0  &  1.0   &-0.30  &  0.24  &  0.46  &  0.31&\\
27&   18  54 47.01&  -30  26  58.2&    2.7&   0.9&   0.0&   0.0&  1.0 &   0.6&  1.7  &  0.7   &0.26  &  0.39  &  0.37  &  0.34&\\
28&   18  54 49.04&  -30  23  32.3&    0.7&   0.5&   0.3&   0.3&  0.0 &   0.0&  0.3  &  0.3   &0.50  &  0.61  & -0.50  &  0.61&\\
29&   18  54 49.07&  -30  23  55.8&    3.7&   1.1&   0.0&   0.0&  2.0 &   0.8&  1.7  &  0.7   &-0.08  &  0.29  &  0.54  &  0.44&\\
30&   18  54 49.13&  -30  25  51.2&    4.7&   1.2&   2.4&   0.9&  2.0 &   0.8&  0.3  &  0.3   &-0.36  &  0.29  & -0.09  &  0.26&\\
31&   18  54 49.22&  -30  25  02.0&   39.1&   3.6&  13.5&   2.1& 23.2 &   2.8&  2.4  &  0.9   &-0.53  &  0.34  &  0.25  &  0.18& 13.6\\
32&   18  54 49.38&  -30  31  05.6&    4.7&   1.2&   0.7&   0.5&  1.0 &   0.6&  3.0  &  1.0   &0.43  &  0.36  &  0.06  &  0.17&\\
33&   18  54 49.88&  -30  31  38.6&    2.4&   0.9&   0.3&   0.3&  1.3 &   0.7&  0.7  &  0.5   &-0.26  &  0.41  &  0.43  &  0.42&\\
34&   18  54 52.97&  -30  35  41.8&    3.7&   1.1&   0.3&   0.3&  1.7 &   0.7&  1.7  &  0.7   &0.00  &  0.00  &  0.38  &  0.31&\\
35&   18  54 54.08&  -30  25  40.1&    3.0&   1.0&   0.0&   0.0&  0.7 &   0.5&  2.4  &  0.9   &0.55  &  0.52  &  0.23  &  0.23&\\
36&   18  54 54.60&  -30  30  42.9&    4.4&   1.2&   0.0&   0.0&  1.7 &   0.7&  2.7  &  0.9   &0.23  &  0.31  &  0.39  &  0.32&\\
37&   18  54 54.72&  -30  26  55.2&    2.7&   0.9&   0.3&   0.3&  0.3 &   0.3&  2.0  &  0.8   &0.65  &  0.54  &  0.00  &  0.00&\\
38&   18  54 54.90&  -30  31  25.4&   11.1&   1.9&   1.3&   0.7&  4.0 &   1.2&  5.7  &  1.4   &0.15  &  0.19  &  0.25  &  0.19&\\
39&   18  54 56.41&  -30  35  08.4&    2.7&   0.9&   2.4&   0.9&  0.3 &   0.3&  0.0  &  0.0   &-0.11  &  0.14  & -0.78  &  0.71& 13.0\\
40&   18  54 57.28&  -30  32  38.1&    6.1&   1.4&   0.0&   0.0&  2.7 &   0.9&  3.4  &  1.1   &0.11  &  0.25  &  0.44  &  0.35&\\
\hline				         						     
\end{tabular}										     
\end{center}										     
\label{sgr_sources}									     
\caption{The X-ray sources found in the {\sl Chandra} observations of
the Sgr field. We show the count rate (per 10ksec) found for each
source in 4 energy bands and the hardness ratios, where HR1=(H-M)/(H+M+S) and 
HR2=(M-S)/(H+M+S), where H is the counts in the 2--8keV band, M the 
counts in the 1--2keV band and S the counts in the 0.3--1keV band. In the 
final column we show the magnitude (close to the $R$ band for that source in the 
UCAC2 catalogue).}
\end{table*}

\begin{table*}
\begin{center}
\begin{tabular}{l@{\hspace{2pt}}c@{\hspace{5pt}}crrrrrrrrrrrr@{\hspace{2pt}}r}
\hline
Src & RA & Dec & \multicolumn{2}{r}{0.3-8keV} & \multicolumn{2}{r}{0.3-1keV} & 
\multicolumn{2}{r}{1-2keV} & \multicolumn{2}{r}{2-8keV} & \multicolumn{2}{c}{HR1} & \multicolumn{2}{c}{HR2} &UCAC2 \\
    & \multicolumn{2}{c}{(2000)} & & $\pm$ &  & $\pm$ & & $\pm$ & & $\pm$ & & $\pm$  & & $\pm$ & mag\\
41&   18  54 57.37&  -30  22  19.9&    2.0&   0.8&   0.0&   0.0&  1.3 &   0.7&  0.7  &  0.5   &-0.30  &  0.48  &  0.65  &  0.58&\\
42&   18  54 59.39&  -30  23  51.6&    3.4&   1.1&   0.0&   0.0&  2.4 &   0.9&  1.0  &  0.6   &-0.41  &  0.43  &  0.71  &  0.57&\\
43&   18  55 00.13&  -30  30  49.5&   17.5&   2.4&   2.7&   0.9&  9.1 &   1.8&  5.7  &  1.4   &-0.19  &  0.17  &  0.37  &  0.25&\\
44&   18  55 04.56&  -30  26  32.5&   17.8&   2.5&   7.1&   1.5&  6.7 &   1.5&  4.0  &  1.2   &-0.15  &  0.14  & -0.02  &  0.12&\\
45&   18  55 05.33&  -30  31  06.2&    2.7&   0.9&   0.7&   0.5&  1.3 &   0.7&  0.7  &  0.5   &-0.22  &  0.34  &  0.22  &  0.34&\\
46&   18  55 06.63&  -30  31  27.3&    2.4&   0.9&   1.0&   0.6&  1.3 &   0.7&  0.0  &  0.0   &-0.56  &  0.50  &  0.13  &  0.41&\\
47&   18  55 06.88&  -30  32  24.3&    1.3&   0.7&   0.3&   0.3&  0.7 &   0.5&  0.3  &  0.3   &-0.31  &  0.48  &  0.31  &  0.48&\\
48&   18  55 07.92&  -30  25  31.7&    4.0&   1.2&   0.0&   0.0&  1.0 &   0.6&  3.0  &  1.0   &0.50  &  0.47  &  0.25  &  0.24&\\
49&   18  55 07.99&  -30  26  20.8&    3.0&   1.0&   0.0&   0.0&  1.3 &   0.7&  1.7  &  0.7   &0.13  &  0.34  &  0.43  &  0.38&\\
50&   18  55 08.38&  -30  30  00.3&    3.4&   1.1&   0.0&   0.0&  1.7 &   0.7&  1.7  &  0.7   &0.00  &  0.00  &  0.50  &  0.41&\\
51&   18  55 10.67&  -30  26  51.1&   44.4&   3.9&   9.8&   1.8& 19.5 &   2.6& 15.2  &  2.3   &-0.10  &  0.10  &  0.22  &  0.15&\\
52&   18  55 10.72&  -30  22  14.1&    9.8&   1.8&   0.0&   0.0&  4.7 &   1.2&  5.1  &  1.3   &0.04  &  0.18  &  0.48  &  0.36&\\
53&   18  55 10.79&  -30  31  25.6&    1.0&   0.6&   0.3&   0.3&  0.0 &   0.0&  0.7  &  0.5   &0.70  &  0.71  & -0.30  &  0.37&\\
54&   18  55 10.90&  -30  21  18.4&    3.0&   1.0&   0.0&   0.0&  0.0 &   0.0&  3.0  &  1.0   &1.00  &  1.05  &  0.00  &  0.00&\\
55&   18  55 11.39&  -30  20  52.4&   12.8&   2.1&   0.7&   0.5&  5.7 &   1.4&  6.4  &  1.5   &0.05  &  0.16  &  0.39  &  0.27&\\
56&   18  55 11.53&  -30  26  19.7&    3.0&   1.0&   0.0&   0.0&  2.0 &   0.8&  1.0  &  0.6   &-0.33  &  0.41  &  0.67  &  0.55&\\
57&   18  55 14.73&  -30  35  41.6&   20.5&   2.6&   9.4&   1.8&  8.8 &   1.7&  2.4  &  0.9   &-0.31  &  0.21  & -0.03  &  0.12&\\
58&   18  55 15.48&  -30  28  09.2&   23.2&   2.8&   0.3&   0.3&  9.1 &   1.8& 13.8  &  2.2   &0.20  &  0.18  &  0.38  &  0.26&\\
59&   18  55 16.15&  -30  28  58.8&    4.0&   1.2&   0.7&   0.5&  2.0 &   0.8&  1.3  &  0.7   &-0.18  &  0.28  &  0.33  &  0.30&\\
60&   18  55 16.79&  -30  31  44.6&    5.4&   1.3&   1.0&   0.6&  2.0 &   0.8&  2.4 &   0.9 &  0.07&   0.23&   0.19&   0.21&\\
61&   18  55 18.42&  -30  33  33.7&    5.4&   1.3&   0.0&   0.0&  0.3 &   0.3&  5.1 &   1.3 &  0.89&   0.78&   0.06&   0.07&\\
62&   18  55 21.81&  -30  27  12.8&    8.1&   1.6&   1.7&   0.7&  3.0 &   1.0&  3.4 &   1.1 &  0.05&   0.19&   0.16&   0.18& 13.6\\
63&   18  55 21.92&  -30  30  10.7&    2.4&   0.9&   1.3&   0.7&  0.3 &   0.3&  0.7 &   0.5 &  0.17&   0.27&  -0.43&   0.42&\\
64&   18  55 22.13&  -30  32  13.1&    8.8&   1.7&   2.7&   0.9&  4.0 &   1.2&  2.0 &   0.8 & -0.23&   0.21&   0.15&   0.19&\\
65&   18  55 22.67&  -30  33  18.5&   13.8&   2.2&   3.0&   1.0&  4.7 &   1.2&  6.1 &   1.4 &  0.10&   0.15&   0.12&   0.13&\\
66&   18  55 24.19&  -30  28  04.2&   11.4&   2.0&   1.3&   0.7&  6.4 &   1.5&  3.7 &   1.1 & -0.24&   0.22&   0.45&   0.31&\\
67&   18  55 25.72&  -30  27  17.0&    2.7&   0.9&   0.0&   0.0&  1.0 &   0.6&  1.7 &   0.7 &  0.26&   0.39&   0.37&   0.34&\\
68&   18  55 26.12&  -30  26  29.0&    8.4&   1.7&   1.3&   0.7&  3.0 &   1.0&  4.0 &   1.2 &  0.12&   0.20&   0.20&   0.19&\\
69&   18  55 27.17&  -30  26  34.1&    3.4&   1.1&   0.7&   0.5&  2.4 &   0.9&  0.3 &   0.3 & -0.62&   0.48&   0.50&   0.44&\\
70&   18  55 27.57&  -30  24  56.1&   10.1&   1.9&   1.3&   0.7&  4.7 &   1.2&  4.0 &   1.2 & -0.07&   0.17&   0.34&   0.24&\\
71&   18  55 27.71&  -30  29  25.5&    9.1&   1.8&   0.3&   0.3&  0.3 &   0.3&  8.4 &   1.7 &  0.90&   0.71&   0.00&   0.00&\\
72&   18  55 27.91&  -30  25  12.0&    3.7&   1.1&   1.3&   0.7&  1.0 &   0.6&  1.3 &   0.7 &  0.08&   0.26&  -0.08&   0.26&\\
73&   18  55 29.27&  -30  25  03.2&    3.4&   1.1&   0.0&   0.0&  0.0 &   0.0&  3.4 &   1.1 &  1.00&   1.05&   0.00&   0.00&\\
74&   18  55 29.79&  -30  28  29.1&    5.7&   1.4&   1.7&   0.7&  2.7 &   0.9&  1.3 &   0.7 & -0.25&   0.25&   0.18&   0.22&\\
75&   18  55 32.01&  -30  34  07.8&    2.7&   0.9&   0.3&   0.3&  1.7 &   0.7&  0.7 &   0.5 & -0.37&   0.39&   0.52&   0.42&\\
76&   18  55 32.64&  -30  27  06.7&    1.0&   0.6&   0.3&   0.3&  0.3 &   0.3&  0.3 &   0.3 &  0.00&   0.00&   0.00&   0.00& 13.6\\
77&   18  55 32.81&  -30  32  58.9&    6.1&   1.4&   1.3&   0.7&  2.4 &   0.9&  2.4 &   0.9 &  0.00&   0.00&   0.18&   0.21& 15.4\\
78&   18  55 33.91&  -30  34  45.4&    2.0&   0.8&   0.7&   0.5&  0.0 &   0.0&  1.3 &   0.7 &  0.65&   0.58&  -0.35&   0.35&\\
79&   18  55 36.65&  -30  29  00.2&    8.1&   1.6&   1.3&   0.7&  3.4 &   1.1&  3.4 &   1.1 &  0.00&   0.00&   0.26&   0.22&\\
80&   18  55 37.27&  -30  30  36.7&  119.5&   6.3&   5.4&   1.3& 52.9 &   4.2& 61.3 &   4.5 &  0.07&   0.07&   0.40&   0.25&\\
\hline
\end{tabular}
\end{center}
\label{sgr_sources}
\caption{Cont.}
\end{table*}



\begin{table*}
\begin{center}
\begin{tabular}{l@{\hspace{2pt}}c@{\hspace{5pt}}crrrrrrrrrrrr@{\hspace{2pt}}r}
\hline
Src & RA & Dec & \multicolumn{2}{r}{0.2-10keV} & \multicolumn{2}{r}{0.2-1keV} &
\multicolumn{2}{r}{1-2keV} & \multicolumn{2}{r}{2-10keV} & \multicolumn{2}{c}{HR1} & \multicolumn{2}{c}{HR2} & UCAC2\\
    & \multicolumn{2}{c}{(2000)} & & $\pm$ &  & $\pm$ & & $\pm$ & & $\pm$ & & $\pm$  & & $\pm$ & mag\\
\hline
 1 & 6  40  14.9 &-50  58  23.1 &   4.2 & 0.8&   0.6 &  0.3 &  1.2 &  0.3 &  2.3 &   0.6 &   0.27 &  0.23&   0.15&   0.14& \\ 
 2 & 6  40  19.8 &-50  58  23.0 &  34.9 & 1.7&  38.6 &  1.8 & 14.5 &  1.1 &  8.7 &   1.2 &  -0.09 &  0.06&  -0.39&   0.23& \\ 
 3 & 6  40  21.5 &-51  00  56.7 &   8.0 & 1.0&   2.3 &  0.5 &  2.8 &  0.4 &  2.9 &   0.6 &   0.01 &  0.09&   0.06&   0.09& \\ 
 4 & 6  40  24.8 &-51  06  16.6 &   5.2 & 1.1&   0.9 &  0.4 &  1.1 &  0.4 &  3.6 &   0.9 &   0.45 &  0.33&   0.04&   0.10& \\ 
 5 & 6  40  28.0 &-50  59  17.6 &   9.2 & 1.0&   8.3 &  0.9 &  2.6 &  0.4 &  6.6 &   1.0 &   0.23 &  0.15&  -0.33&   0.21& \\ 
 6 & 6  40  28.6 &-50  57  02.3 &   3.3 & 0.7&   0.9 &  0.3 &  1.1 &  0.3 &  1.1 &   0.5 &   0.00 &  0.00&   0.06&   0.14& \\  
 7 & 6  40  34.1 &-50  59  39.8 &   6.7 & 0.8&   9.4 &  0.8 &  1.4 &  0.4 &  0.5 &   0.5 &  -0.08 &  0.07&  -0.71&   0.43& 11.4\\ 
 8 & 6  40  36.4 &-50  53  18.5 &   4.6 & 0.7&   1.2 &  0.3 &  1.7 &  0.4 &  1.5 &   0.5 &  -0.05 &  0.15&   0.11&   0.13& \\ 
 9 & 6  40  36.4 &-50  59  17.1 &   4.3 & 0.5&   2.0 &  0.4 &  1.7 &  0.4 &  3.0 &   0.7 &   0.19 &  0.17&  -0.04&   0.09& \\  
10 & 6  40  36.7 &-50  49  20.7 &   3.5 & 0.9&   0.6 &  0.3 &  1.4 &  0.4 &  1.2 &   0.6 &  -0.06 &  0.23&   0.25&   0.22& \\  
11 & 6  40  41.2 &-50  47  49.7 &   4.8 & 0.9&   2.3 &  0.4 &  1.4 &  0.4 &  0.7 &   0.6 &  -0.16 &  0.19&  -0.20&   0.18& \\ 
12 & 6  40  44.7 &-51  06  22.0 &   7.3 & 1.0&   1.9 &  0.5 &  2.9 &  0.5 &  2.6 &   0.6 &  -0.04 &  0.11&   0.14&   0.12& \\ 
13 & 6  40  50.0 &-50  56  31.8 &   2.8 & 0.4&   2.4 &  0.4 &  0.9 &  0.3 &  1.3 &   0.6 &   0.09 &  0.15&  -0.33&   0.22& \\ 
14 & 6  40  53.6 &-50  51  20.6 &   2.2 & 0.6&   0.6 &  0.3 &  0.8 &  0.2 &  0.7 &   0.4 &  -0.05 &  0.21&   0.10&   0.18& \\ 
15 & 6  40  56.1 &-50  52  32.6 &   2.6 & 0.5&   6.1 &  0.7 &  0.8 &  0.3 &  0.0 &   0.2 &  -0.12 &  0.09&  -0.77&   0.52& 12.2\\ 
16 & 6  40  59.2 &-50  55  51.1 &   6.2 & 0.6&   5.9 &  0.6 &  2.2 &  0.3 &  1.5 &   0.5 &  -0.07 &  0.07&  -0.39&   0.24& \\ 
17 & 6  41  00.2 &-50  57  14.3 &   2.8 & 0.5&   1.7 &  0.2 &  0.8 &  0.2 &  0.0 &   0.1 &  -0.32 &  0.21&  -0.36&   0.24& \\ 
18 & 6  41  00.6 &-50  47  01.3 &   4.1 & 0.8&   1.6 &  0.4 &  1.3 &  0.4 &  1.0 &   0.5 &  -0.08 &  0.17&  -0.08&   0.15& \\ 
19 & 6  41  04.6 &-51  00  00.0 &   7.7 & 0.6&  12.7 &  0.8 &  2.3 &  0.3 &  1.6 &   0.5 &  -0.04 &  0.04&  -0.63&   0.39& 15.4\\ 
20 & 6  41  06.9 &-50  47  40.8 &   3.1 & 0.7&   3.1 &  0.6 &  1.7 &  0.5 &  2.2 &   0.9 &   0.07 &  0.15&  -0.20&   0.16& \\ 
21 & 6  41  07.5 &-51  07  05.5 &   4.4 & 0.8&   2.9 &  0.5 &  4.0 &  0.6 &  8.9 &   1.2 &   0.31 &  0.21&   0.07&   0.07& \\ 
22 & 6  41  09.1 &-51  02  06.7 &   3.8 & 0.5&   3.6 &  0.5 &  1.9 &  0.3 &  2.3 &   0.6 &   0.05 &  0.09&  -0.22&   0.15& \\ 
23 & 6  41  09.6 &-51  09  07.8 &  12.2 & 1.2&   9.0 &  1.0 &  2.6 &  0.5 &  9.1 &   1.4 &   0.31 &  0.21&  -0.31&   0.20& \\ 
24 & 6  41  12.3 &-50  54  20.3 &   3.0 & 0.5&   2.6 &  0.4 &  2.2 &  0.3 &  1.6 &   0.5 &  -0.09 &  0.11&  -0.06&   0.09& \\ 
25 & 6  41  18.9 &-50  54  48.8 &   2.2 & 0.4&   2.6 &  0.4 &  1.3 &  0.2 &  0.4 &   0.4 &  -0.21 &  0.16&  -0.30&   0.21& \\ 
26 & 6  41  22.1 &-51  04  11.7 &   5.0 & 0.6&   0.8 &  0.2 &  2.2 &  0.3 &  2.0 &   0.4 &  -0.04 &  0.10&   0.28&   0.18& \\ 
27 & 6  41  23.9 &-50  47  58.7 &   5.6 & 0.8&   4.1 &  0.6 &  3.5 &  0.5 &  3.1 &   0.8 &  -0.04 &  0.09&  -0.06&   0.08& \\ 
28 & 6  41  28.6 &-51  00  58.0 &   2.9 & 0.4&   0.1 &  0.1 &  0.8 &  0.2 &  1.9 &   0.3 &   0.39 &  0.28&   0.25&   0.18& \\ 
29 & 6  41  30.8 &-50  56  59.1 &   4.8 & 0.5&   9.5 &  0.7 &  1.4 &  0.3 &  0.6 &   0.4 &  -0.07 &  0.06&  -0.70&   0.44& 12.1\\ 
30 & 6  41  33.0 &-50  59  34.1 &   2.1 & 0.4&   2.1 &  0.3 &  0.8 &  0.2 &  0.1 &   0.3 &  -0.23 &  0.18&  -0.43&   0.28& \\ 
31 & 6  41  34.5 &-50  48  24.9 &   7.8 & 0.6&   4.1 &  0.6 &  3.4 &  0.5 &  4.4 &   0.8 &   0.08 &  0.09&  -0.06&   0.07& \\ 
32 & 6  41  37.7 &-50  56  47.7 &   3.7 & 0.5&   1.4 &  0.2 &  1.2 &  0.2 &  0.7 &   0.3 &  -0.15 &  0.14&  -0.06&   0.09& \\ 
33 & 6  41  40.3 &-51  00  05.4 &   7.5 & 0.6&   6.4 &  0.5 &  1.1 &  0.2 &  0.0 &   0.1 &  -0.15 &  0.10&  -0.71&   0.49& 9.2\\ 
34 & 6  41  40.7 &-51  04  47.1 &   2.1 & 0.4&   1.9 &  0.4 &  0.6 &  0.3 &  1.1 &   0.6 &   0.14 &  0.20&  -0.36&   0.26& \\ 
35 & 6  41  44.3 &-50  51  54.0 &   2.5 & 0.5&   2.0 &  0.4 &  2.0 &  0.4 &  3.4 &   0.6 &   0.19 &  0.15&   0.00&   0.00& \\ 
36 & 6  41  46.1 &-50  50  22.9 &  65.8 & 3.9&  68.5 &  1.9 & 29.6 &  1.2 & 26.9 &   1.4 &  -0.02 &  0.02&  -0.31&   0.18& \\ 
37 & 6  41  46.4 &-50  48  03.0 &   4.8 & 0.7&   2.3 &  0.4 &  1.2 &  0.3 &  1.1 &   0.4 &  -0.02 &  0.11&  -0.24&   0.18& \\ 
38 & 6  41  48.0 &-50  49  40.3 &   4.9 & 0.8&   4.3 &  0.6 &  2.5 &  0.4 &  7.0 &   0.9 &   0.33 &  0.21&  -0.13&   0.09& \\ 
39 & 6  41  48.3 &-51  13  43.2 &  37.6 & 1.7&   2.3 &  0.9 & 22.1 &  1.8 & 53.9 &   3.4 &   0.41 &  0.27&   0.25&   0.17& \\ 
40 & 6  41  48.9 &-50  59  41.9 &   1.6 & 0.4&   0.4 &  0.2 &  0.7 &  0.2 &  0.4 &   0.2 &  -0.20 &  0.22&   0.20&   0.22& \\ 
41 & 6  41  49.2 &-50  47  11.5 &  10.1 & 1.0&   5.0 &  0.7 &  4.9 &  0.6 & 13.2 &   1.4 &   0.36 &  0.23&   0.00&   0.04& \\ 
42 & 6  41  49.2 &-51  06  51.2 &   3.1 & 0.7&   0.5 &  0.4 &  1.4 &  0.4 &  5.4 &   1.0 &   0.55 &  0.38&   0.12&   0.11& \\ 
43 & 6  41  51.5 &-50  57  14.9 &   9.5 & 0.7&  11.0 &  0.7 &  4.4 &  0.5 &  5.0 &   0.6 &   0.03 &  0.04&  -0.32&   0.19& \\ 
44 & 6  41  51.6 &-50  46  38.6 &   3.5 & 0.7&   5.0 &  0.7 &  3.0 &  0.5 &  7.2 &   1.1 &   0.28 &  0.19&  -0.13&   0.10& \\ 
45 & 6  41  59.5 &-51  09  46.8 &   5.8 & 1.1&   2.8 &  0.6 &  1.4 &  0.4 &  0.2 &   0.4 &  -0.27 &  0.21&  -0.32&   0.25& 16.2\\ 
46 & 6  42  00.2 &-51  00  29.0 &   5.6 & 0.6&   4.5 &  0.5 &  3.1 &  0.4 &  2.9 &   0.6 &  -0.02 &  0.07&  -0.13&   0.10& \\ 
47 & 6  42  05.3 &-51  03  44.8 &   2.4 & 0.6&   2.0 &  0.4 &  1.1 &  0.4 &  1.2 &   0.7 &   0.02 &  0.19&  -0.21&   0.18& \\ 
48 & 6  42  05.8 &-51  07  47.4 &   6.4 & 1.0&   4.4 &  0.8 &  3.1 &  0.6 &  4.6 &   1.1 &   0.12 &  0.13&  -0.11&   0.10& \\ 
49 & 6  42  09.4 &-51  02  51.6 &   4.3 & 0.7&   1.0 &  0.3 &  1.1 &  0.3 &  2.1 &   0.4 &   0.24 &  0.18&   0.02&   0.10& \\ 
50 & 6  42  11.9 &-50  53  37.6 &   8.2 & 0.8&   2.6 &  0.3 &  2.5 &  0.3 &  3.0 &   0.5 &   0.06 &  0.08&  -0.01&   0.05& \\ 
51 & 6  42  17.4 &-51  01  44.4 &  10.9 & 0.9&   2.8 &  0.4 &  3.8 &  0.5 &  4.0 &   0.6 &   0.02 &  0.07&   0.09&   0.08& \\ 
52 & 6  42  25.0 &-50  59  52.3 &   5.0 & 0.6&   4.3 &  0.6 &  2.4 &  0.4 &  3.3 &   0.8 &   0.09 &  0.10&  -0.19&   0.13& \\ 
53 & 6  42  47.4 &-51  00  25.7 &  21.5 & 1.7&  15.4 &  1.4 &  7.9 &  1.0 &  9.7 &   1.5 &   0.05 &  0.06&  -0.23&   0.14& \\ 
\hline
\end{tabular}
\end{center}
\label{carina_sources}
\caption{The X-ray sources in the Carina field as detected using {sl
XMM-Newton}. We show the count rates in Cts/ksec in the 0.2-10keV, 0.2-1keV, 1-2keV and 2-10keV 
energy bands. The count rates refer to that detected in the EPIC MOS(1+2) detector, or the 
equivalent if detected only in the EPIC pn detector. We also show the hardness
ratios, where HR1=(H-M)/(H+M+S) and
HR2=(M-S)/(H+M+S), where H is the counts in the 2--10keV band, M the
counts in the 1--2keV band and S the counts in the 0.2--1keV band. In the
final column we show the magnitude (close to the $R$ band for that source in the
UCAC2 catalogue).}
\end{table*}
\newpage


\end{document}